\documentclass[11pt]{article}
\PassOptionsToPackage{table}{xcolor}
\usepackage[T1]{fontenc}
\usepackage{chao}
\usepackage{mathtools}
\usepackage{enumitem}
\usepackage{booktabs,subcaption,colortbl,hhline,tabularx}
\usepackage[capitalise,nameinlink,noabbrev]{cleveref}
\crefname{lemma}{Lemma}{Lemmas}
\crefname{theorem}{Theorem}{Theorems}
\date{}
\hypersetup{pdfauthor={Yuhao Liu, Ke Shi, Chao Xu},pdftitle={A Polynomial Time Algorithm for Finding a Minimum 5-Partition of a Submodular Function}}

\providecommand{\green}{green!10}
\providecommand{\red}{red!10}
\providecommand{\yellow}{yellow!10}
\providecommand{\blue}{blue!10}

\newcommand{\compat}{\lhd}
\title{A Polynomial Time Algorithm for Finding a Minimum
5-Partition of a Submodular Function}
\newcommand{\authoremail}[1]{\href{mailto:#1}{\textcolor{blue}{#1}}}
\author{\large Yuhao Liu\thanks{\authoremail{yuhaoliu@cs.wisc.edu}, University of Wisconsin-Madison.}
\hspace{2em}Ke Shi\thanks{\authoremail{self.ke.shi@gmail.com}, University of Science and Technology of China.}
\hspace{2em}Chao Xu\thanks{\authoremail{the.chao.xu@gmail.com}, University of Electronic Science and Technology of China.}}
\begin{document}
\makeatletter
\begingroup
\def\@fnsymbol#1{\ensuremath{\ifcase#1\or\dagger\or\ddagger\or\mathsection\else\@ctrerr\fi}}
\maketitle
\endgroup
\makeatother

\begin{abstract}
Given a real-valued submodular function on a finite ground set, the minimum $k$-partition problem asks for a partition into $k$ nonempty parts minimizing the sum of their function values. We give a polynomial-time algorithm for $k=5$. The idea follows directly from the previous $k=4$ result, where there is precisely one obstruction for generalizing it to $k=5$. We show the special case can be handled by reasoning about a restricted family of $3$-partitions. 
\par\smallskip\noindent\textbf{Keywords:} Submodular functions; Partitioning; Exact algorithms.
\end{abstract}

\section{Introduction}

Let $V$ be a finite set and let $f:2^V\to\mathbb R$ be \emph{submodular},
meaning that $f(X)+f(Y)\ge f(X\cap Y)+f(X\cup Y)$ for all $X,Y\subseteq V$.
The minimum $k$-partition problem asks for a partition
$\mathcal P=\{V_1,\ldots,V_k\}$ of $V$ into nonempty parts minimizing
$f(\mathcal P)=\sum_{i=1}^k f(V_i)$.
The problem is NP-hard when $k$ is part of the input~\cite{GH}, and
polynomial-time algorithms were known for $k=2$~\cite{Queyranne},
$k=3$~\cite{OFN}, and $k=4$~\cite[Theorem~6.1]{HLMSX}.
Our algorithm extends the compatibility approach of Hirayama
et al.~\cite{HLMSX} to $k=5$.

\begin{theorem}\label{thm:main}
For a submodular function $f:2^V\to\mathbb R$ with $|V|\ge5$ given by a value
oracle, a minimum $5$-partition can be found in polynomial time.
\end{theorem}

By Chandrasekaran and Chekuri~\cite[Corollary~4.1]{CC}, this also gives
a polynomial-time algorithm for symmetric submodular 6-partition,
including minimum-sum 6-partition in hypergraphs with nonnegative
edge weights.

After handling small parts, the 4-partition algorithm of Hirayama
et al.~\cite{HLMSX} contracts each pair of parts of a minimum non-trivial
3-partition, where non-trivial means having at least two parts of size at
least $2$. Their compatibility
theorem guarantees that contracting some pair preserves a minimum
4-partition.
The analogous compatibility condition for $k=5$ would require three
parts of a 4-partition to be contained in parts of a minimum
5-partition. Their concluding example, whose pattern of nonempty pairwise
intersections appears in \cref{fig:canonical-configuration}, shows
that this can fail even for graphs~\cite[Section~7]{HLMSX}.

\cref{lem:canonical-short} proves that every incompatible configuration of a
minimum non-trivial 4-partition with the minimum 5-partition has this pattern
of nonempty cells when the minimum 5-partition is unique and all its parts
have size at least $5$. We overcome this obstruction by finding a minimum
3-partition $\{P,Q,R\}$ among those in which $P$ and $Q$ lie in the same part
of the 4-partition and at least two parts have size at least $3$.
\cref{lem:noncrossing} shows that $P$ and $Q$ each lie in a part of the
minimum 5-partition, making $\{P,Q,R\}$ compatible with it. Contracting $P$
and $Q$ therefore preserves the minimum 5-partition even when the 4-partition
is incompatible. It also leaves the updated 4-partition a minimum non-trivial
4-partition, so the algorithm can repeat this step.

\section{Preliminaries}
\label{sec:preliminaries}

We follow the terminology and notation of Hirayama et al.~\cite{HLMSX}.
We work in the value-oracle model, where evaluations of $f$ and elementary
arithmetic operations and comparisons take unit time.
All partitions have nonempty parts, and $f(\mathcal P)$ denotes the sum
of $f$ over the parts of $\mathcal P$. For $U\subseteq V$, let $f_{\setminus U}$ be
the restriction of $f$ to $V\setminus U$. For nonempty $U\subseteq V$, let
$f_{/U}$ be the function on $(V\setminus U)\cup\{u\}$ obtained from $f$ by
\emph{contracting} $U$ into a new element $u\notin V$, defined by
$f_{/U}(S)=f(S)$ if $u\notin S$ and $f_{/U}(S)=f((S\setminus\{u\})\cup U)$ if
$u\in S$. Both operations preserve submodularity. For a family $\mathcal A$
of nonempty disjoint subsets of $V$, let $f_{/\mathcal A}$ be obtained by
contracting each $A\in\mathcal A$ into a distinct new element $u_A$.
The \emph{lift} of a subset $S$ of the contracted ground set replaces each
$u_A\in S$ by the elements of $A$, and the \emph{lift}
$\operatorname{lift}(\mathcal P)$ of a partition $\mathcal P$ of the
contracted ground set is the partition of $V$ formed by the lifts of its
parts. Lifting preserves partition values, maps distinct partitions to
distinct lifts, and composes across iterated contractions. Unless stated
otherwise, cardinalities refer to the current ground set.

A set is \emph{noncrossing} with a partition if it is contained in one
of its parts, and otherwise \emph{crosses} the partition.
A partition $\mathcal X$ is \emph{compatible} with
$\mathcal Y$, written $\mathcal X\compat\mathcal Y$, if at least
$|\mathcal X|-1$ parts of $\mathcal X$ are noncrossing with $\mathcal Y$,
and is \emph{incompatible} with $\mathcal Y$ otherwise.
The noncrossing parts need not lie in distinct parts of $\mathcal Y$.
For an integer $\ell\ge2$, a partition is $\ell$-\emph{non-trivial}
if at least two parts have size at least $\ell$.
A partition is \emph{non-trivial} if it is $2$-non-trivial, and
\emph{trivial} otherwise.
For an integer $h\ge1$, a partition is $h$-\emph{size} if every part has size
at least $h$.

For partitions $\mathcal X=\{X_1,\ldots,X_p\}$ and
$\mathcal Y=\{Y_1,\ldots,Y_q\}$, write $Z_{ij}=X_i\cap Y_j$.
Their \emph{configuration} is the matrix $M$ with $M_{ij}=|Z_{ij}|$.
The sets $Z_{ij}$ are its \emph{cells}. Configurations are equivalent
up to row and column permutations. Rows and columns also denote
the corresponding parts $X_1,\ldots,X_p$ and $Y_1,\ldots,Y_q$.
The \emph{configuration graph} $G(M)$ is the bipartite graph with a row vertex
$X_i$ for each row, a column vertex $Y_j$ for each column, and an edge $X_iY_j$
whenever $M_{ij}>0$. For a graph $G$ and a vertex set $U$ of $G$, write $G[U]$
for the subgraph induced by $U$. For a vertex set $U$ of $G(M)$, a cell
$Z_{ij}$ is \emph{internal to $U$} if both $X_i$ and $Y_j$ belong to $U$.

\subsection{Optimization and Tie-Breaking}
\label{sec:primitives}

For disjoint $I,E\subseteq V$, minimizing $X\mapsto f(X)+f(V\setminus X)$
over $I\subseteq X\subseteq V\setminus E$ has only the candidate $X=I$ when
$I\cup E=V$, and otherwise reduces to submodular function minimization (SFM)
on $V\setminus(I\cup E)$ via $A\mapsto f(I\cup A)+f(V\setminus(I\cup A))$.
Enumerating disjoint $I,E$ of size $h$ therefore finds a minimum $h$-size
2-partition for every fixed $h\ge2$ in polynomial time~\cite[Lemma~5.3]{HLMSX}.
For $|V|\ge2$, a minimum 2-partition is found by minimizing the symmetric
submodular function $S\mapsto f(S)+f(V\setminus S)$ over nonempty
$S\subsetneq V$ using Queyranne's symmetric submodular function minimization
(SSFM) algorithm~\cite{Queyranne}.
For fixed $\ell\ge2$, a minimum $\ell$-non-trivial 3-partition
(called $(1,\ell)$-size in~\cite{HLMSX}) can be found in polynomial time
by~\cite[Lemmas~5.1 and~5.4]{HLMSX}. For $|V|\ge18\ell-16$, the small-part
enumeration in~\cite[Lemma~5.4]{HLMSX} pairs each enumerated set $S$ with
$1\le|S|<\ell$ with a minimum $\ell$-size 2-partition of $f_{\setminus S}$,
and each $S$ with $\ell\le|S|\le4\ell-4$ with a minimum 2-partition of
$f_{\setminus S}$, which has a part of size at least $\ell$ because
$|V\setminus S|\ge14\ell-12\ge2\ell-1$.
A minimum 3-partition~\cite{OFN} (see~\cite[Section~3]{HLMSX}) and a minimum
4-partition~\cite[Theorem~6.1]{HLMSX} (using $\ell=2$ above) can likewise be
found in polynomial time. Infeasible subproblems contribute no candidate.

Let $V_0$ be the initial ground set of size $n_0=|V_0|$, number the unordered
pairs $e\in\binom{V_0}{2}$ by $r(e)\in\{1,\ldots,\binom{n_0}{2}\}$, and define
$\psi(S)=-\sum_{e\in\binom{S}{2}}2^{r(e)}$.
The function $\psi$ is submodular, and $\psi(\mathcal P)$ determines the
unordered pairs inside parts of $\mathcal P$, hence $\mathcal P$.
Throughout, compare values lexicographically as $(f,\psi)$, passing to
$(f_{\setminus U},\psi_{\setminus U})$ and $(f_{/U},\psi_{/U})$ under
restriction and contraction. Every resulting pair $(g,\phi)$ orders
subsets and partitions identically to $g+\varepsilon\phi$ for sufficiently
small $\varepsilon>0$, so minimum partitions are unique and submodular
structural lemmas apply. Applying the SFM and SSFM reductions above to
$(g,\phi)$ gives submodular and symmetric submodular pairs $(f',\psi')$.

Queyranne's SSFM algorithm~\cite{Queyranne} and the strongly polynomial,
fully combinatorial SFM algorithm at the end of Section~7 of Iwata and
Orlin~\cite{IO} apply directly to every symmetric submodular or submodular
pair $(f',\psi')$ constructed above. On an $m$-element ground set, Queyranne's
algorithm uses only objective evaluations, additions, subtractions,
comparisons, and contractions, while Iwata and Orlin's algorithm keeps the
fixed denominator $16m^4$ for convex combinations and uses only additional
multiplications by polynomially bounded integers, rounding bounded ratios via
comparisons with integer multiples of the divisor without dividing by function
values. Their executions on $(f',\psi')$ agree with those on
$f'+\varepsilon\psi'$ for sufficiently small $\varepsilon>0$, preserving
optimality and the strongly polynomial operation bounds. Since $\psi'$ takes
$O(n_0^2)$-bit integer values and all integer multipliers and denominators
are polynomially bounded across polynomially many additions, the second
coordinates have bit length polynomial in $n_0$.
Any algorithm that accesses its input only through restrictions, contractions,
additions, comparisons, and these SFM and SSFM primitives therefore applies
directly to $(g,\phi)$ and agrees with its execution on $g+\varepsilon\phi$
for sufficiently small $\varepsilon>0$.

\subsection{The Auxiliary 3-Partition}

For a 4-partition $\mathcal X$ of $V$, let $\mathcal F_{\mathcal X}$ be the
family of $3$-non-trivial 3-partitions $\mathcal Z=\{P,Q,R\}$ of $V$ with $P$
and $Q$ contained in the same part of $\mathcal X$. Only $P$ and $Q$ are
noncrossing with $\mathcal X$, since $R$ contains the other three parts of
$\mathcal X$.

\begin{lemma}\label{lem:auxiliary-three}
For a submodular function $f:2^V\to\mathbb R$ and a $4$-partition $\mathcal X$
of $V$ with $\mathcal F_{\mathcal X}\ne\varnothing$, a minimum partition in
$\mathcal F_{\mathcal X}$ can be found in polynomial time.
\end{lemma}
\begin{proof}
Fix $X_i\in\mathcal X$ and let $g=f_{/(V\setminus X_i)}$, with new element $z$.
Compute a minimum $3$-non-trivial 3-partition of $g$ as in
\cref{sec:primitives}.
Its lift belongs to $\mathcal F_{\mathcal X}$, since the lift of the part
containing $z$ contains $V\setminus X_i$ and therefore has size at least $3$,
while another part already has size at least $3$. Conversely, for any
$\{P,Q,R\}\in\mathcal F_{\mathcal X}$ with $P,Q\subseteq X_i$ and
$|R\cap X_i|\ge2$, the corresponding partition
$\{P,Q,(R\cap X_i)\cup\{z\}\}$ of $g$ is $3$-non-trivial and is covered by
this call.

Otherwise $|R\cap X_i|\le1$, so $R=V\setminus X_i$ or
$R=(V\setminus X_i)\cup\{v\}$ for some $v\in X_i$. For each such $R$,
compute a minimum 2-partition of $f_{\setminus R}$ when $|X_i\setminus R|\ge5$,
enumerate all 2-partitions $\{P,Q\}$ of $X_i\setminus R$ with
$\max(|P|,|Q|)=3$ when $|X_i\setminus R|=4$, and skip $|X_i\setminus R|\le3$.
Taking a minimum-value candidate over $X_i\in\mathcal X$ proves the lemma.
\end{proof}


\subsection{Uncrossing}

\begin{lemma}\label{lem:uncrossing}
Let $f:2^V\to\mathbb R$ be submodular, let $\mathcal X$ and $\mathcal Y$ be
partitions of $V$, and let $U$ be a vertex set of their configuration graph
$G(M)$ with $|U|\ge2$ such that $G(M)[U]$ is connected. For each $v\in U$, let
$S_v$ be the part of $\mathcal X$ or $\mathcal Y$ corresponding to $v$, and
write $U^*=\bigcup_{v\in U}S_v$. There is a partition $\mathcal L_U$ of the
union of the cells internal to $U$ into $|U|-1$ parts such that
$\sum_{v\in U}f(S_v)\ge f(U^*)+f(\mathcal L_U)$.
\end{lemma}
\begin{proof}
Let $m=|U|$. Since $G(M)[U]$ is connected, order its vertices as
$v_1,\ldots,v_m$ so that each vertex after the first has an
earlier neighbor. Set $A_i=\bigcup_{r=1}^i S_{v_r}$ for $i=1,\ldots,m$
and $L_i=A_{i-1}\cap S_{v_i}$ for $i=2,\ldots,m$.
Each $L_i$ is nonempty because $v_i$ has an earlier neighbor.
The sets $L_i$ are disjoint: a nonempty cell internal to $U$ is contained in
$L_i$ if and only if $v_i$ is the later of its row and column in the ordering.
Hence $\mathcal L_U=\{L_2,\ldots,L_m\}$ partitions the union of the
cells internal to $U$ into $m-1=|U|-1$ parts.
By submodularity, $f(A_{i-1})+f(S_{v_i})\ge f(A_i)+f(L_i)$ for $i=2,\ldots,m$.
Summing over $i=2,\ldots,m$ and using $A_1=S_{v_1}$ and $A_m=U^*$ gives the
desired inequality.
\end{proof}

\section{Minimum Non-trivial 4-Partitions}
\begin{lemma}\label{lem:nontrivial-four}
For a submodular function $f:2^V\to\mathbb R$ with $|V|\ge6$ given by a value
oracle, a minimum non-trivial $4$-partition can be found in polynomial time.
\end{lemma}
\begin{proof}
We use the exchanges of Hirayama et al.~\cite[Section~4]{HLMSX},
checking the additional non-triviality condition on the 4-partition.
Suppose every minimum non-trivial 4-partition is $13$-size.
Fix a minimum non-trivial 4-partition $\mathcal Y=\{Y_1,Y_2,Y_3,Y_4\}$
and a minimum $5$-non-trivial 3-partition $\mathcal X=\{X_1,X_2,X_3\}$.
We show that $\mathcal X\compat\mathcal Y'$ for some minimum non-trivial
4-partition $\mathcal Y'$.

Let $M$ be the configuration of $\mathcal X$ and $\mathcal Y$, with
$\mathcal X$ as rows and $\mathcal Y$ as columns, and consider vertex degrees
in $G(M)$. Call a row \emph{full} if all its cells are nonempty,
and a cell \emph{heavy} if its size is at least $5$.
Every column has a heavy cell, since $13>3\cdot4$.
At least two rows have size at least $5$.
The cited exchanges, obtained by \cref{lem:uncrossing}, satisfy
$f(\mathcal X')+f(\mathcal Y')\le f(\mathcal X)+f(\mathcal Y)$.
Whenever $\mathcal X'$ is $5$-non-trivial and $\mathcal Y'$ is non-trivial,
both are optimal~\cite[Proposition~4.1]{HLMSX}.
Thus it suffices that $\mathcal X\compat\mathcal Y'$, or that $\mathcal Y'$ has
a part of size at most $12$, contradicting the hypothesis.

Call the configuration a \emph{cross} at column $Y_j$ if, after
relabeling the rows and setting $\{j,k,l,m\}=\{1,2,3,4\}$, the cells
$Z_{1j},Z_{2j}$ are nonempty with $Z_{1j}$ heavy, and $Z_{3k},Z_{3l},Z_{3m}$
are nonempty with $|Z_{3k}|\ge2$ (\cref{fig:cross-exchange}).
The cross exchange of~\cite[Lemma~4.1]{HLMSX} sets
$\mathcal X'=\{Z_{1j},Z_{2j},X_3\cup\bigcup_{t\ne j}Y_t\}$ and
$\mathcal Y'=\{X_1\cup X_2\cup Y_j,Z_{3k},Z_{3l},Z_{3m}\}$.
The heavy cell $Z_{1j}$ and $X_3\cup\bigcup_{t\ne j}Y_t$ ensure that
$\mathcal X'$ is $5$-non-trivial, while $Z_{3k}$ and $X_1\cup X_2\cup Y_j$
make $\mathcal Y'$ non-trivial. The rows $X_1$ and $X_2$ are noncrossing with
$\mathcal Y'$, giving compatibility. Hence assume that no cross exists.

\begin{figure}[htbp]
\centering
\begingroup
\setlength{\tabcolsep}{5pt}
\renewcommand{\arraystretch}{1.05}
\begin{subfigure}[b]{.32\textwidth}
\centering
\begin{tabular}{c|cccc|}
\multicolumn{1}{c}{} & \multicolumn{1}{c}{$Y_j$} & \multicolumn{1}{c}{$Y_k$}
  & \multicolumn{1}{c}{$Y_l$} & \multicolumn{1}{c}{$Y_m$} \\
\hhline{~|----|}
$X_1$ & $5^+$ & $?$ & $?$ & $?$ \\
$X_2$ & $1^+$ & $?$ & $?$ & $?$ \\
$X_3$ & $?$ & $2^+$ & $1^+$ & $1^+$ \\
\hhline{~|----|}
\end{tabular}
\caption{$M$}
\end{subfigure}\hfill
\begin{subfigure}[b]{.32\textwidth}
\centering
\arrayrulecolor{black}
\begin{tabular}{c|cccc|}
\multicolumn{1}{c}{} & \multicolumn{1}{c}{$Y_j$} & \multicolumn{1}{c}{$Y_k$}
  & \multicolumn{1}{c}{$Y_l$} & \multicolumn{1}{c}{$Y_m$} \\
\hhline{~|----|}
$X_1$ & \cellcolor{\blue}$5^+$ & \cellcolor{\blue}$?$ & \cellcolor{\blue}$?$ & \cellcolor{\blue}$?$ \\
$X_2$ & \cellcolor{\blue}$1^+$ & \cellcolor{\blue}$?$ & \cellcolor{\blue}$?$ & \cellcolor{\blue}$?$ \\
\hhline{~|>{\arrayrulecolor{\blue}}->{\arrayrulecolor{black}}---|}
$X_3$ & \multicolumn{1}{|c|}{\cellcolor{\blue}$?$}
    & \multicolumn{1}{c|}{\cellcolor{\yellow}$2^+$}
    & \multicolumn{1}{c|}{\cellcolor{\green}$1^+$}
    & \cellcolor{\red}$1^+$ \\
\hhline{~|----|}
\end{tabular}
\caption{$\mathcal Y'$}
\end{subfigure}\hfill
\begin{subfigure}[b]{.32\textwidth}
\centering
\arrayrulecolor{black}
\begin{tabular}{c|c|ccc|}
\multicolumn{1}{c}{} & \multicolumn{1}{c}{$Y_j$} & \multicolumn{1}{c}{$Y_k$}
  & \multicolumn{1}{c}{$Y_l$} & \multicolumn{1}{c}{$Y_m$} \\
\hhline{~|----|}
$X_1$ & \cellcolor{\red}$5^+$ & \cellcolor{\blue}$?$ & \cellcolor{\blue}$?$ & \cellcolor{\blue}$?$ \\
\hhline{~|->{\arrayrulecolor{\blue}}--->{\arrayrulecolor{black}}|}
$X_2$ & \cellcolor{\green}$1^+$ & \cellcolor{\blue}$?$ & \cellcolor{\blue}$?$ & \cellcolor{\blue}$?$ \\
\hhline{~|->{\arrayrulecolor{\blue}}--->{\arrayrulecolor{black}}|}
$X_3$ & \multicolumn{1}{|c}{\cellcolor{\blue}$?$}
    & \cellcolor{\blue}$2^+$ & \cellcolor{\blue}$1^+$ & \cellcolor{\blue}$1^+$ \\
\hhline{~|----|}
\end{tabular}
\caption{$\mathcal X'$}
\end{subfigure}
\endgroup
\caption{The cross exchange, where $\{j,k,l,m\}=\{1,2,3,4\}$.
Here $a^+$ denotes an entry at least $a$, and $?$ denotes an arbitrary entry.
Colors mark the parts of
$\mathcal Y'=\{X_1\cup X_2\cup Y_j,Z_{3k},Z_{3l},Z_{3m}\}$ and
$\mathcal X'=\{Z_{1j},Z_{2j},X_3\cup\bigcup_{t\ne j}Y_t\}$.
After relabeling, $|Z_{1j}|\ge5$ and $|Z_{3k}|\ge2$.
If all columns have size at least $13$, then $\mathcal X'$ is
$5$-non-trivial and $\mathcal Y'$ is non-trivial.}
\label{fig:cross-exchange}
\end{figure}


If two rows have degree $1$, compatibility already holds.
If a row has degree $2$, relabel it as $X_1$. If it meets columns $Y_j$ and
$Y_k$, use the exchange of~\cite[Lemma~4.2]{HLMSX}:
\[
 \mathcal X'=\{X_1\cup Y_j\cup Y_k,Y_l,Y_m\},\qquad
 \mathcal Y'=\{Z_{1j},Z_{1k},X_2,X_3\},
\]
where $\{j,k,l,m\}=\{1,2,3,4\}$. The parts $Y_l$ and $Y_m$ ensure that
$\mathcal X'$ is $5$-non-trivial. If $|X_1|<5$, both $X_2$ and $X_3$ have size
at least $5$. Otherwise $|X_1|\ge5$, so one of $Z_{1j},Z_{1k}$ has size at
least $3$, while at least one of $X_2,X_3$ has size at least $5$. Thus
$\mathcal Y'$ is non-trivial, and $\mathcal X\compat\mathcal Y'$ since
$X_2,X_3\in\mathcal Y'$. Assume henceforth that every row has degree $1$ or at
least $3$, and at most one row has degree $1$.

\emph{Case 1: every row has degree at least $3$.}
Follow the column-degree cases
of~\cite[Theorem~4.1]{HLMSX}.

When every column has degree at least $2$, any row with an empty cell has
size at most $3$: such a row of size at least $4$ has three nonempty cells,
including one of size at least $2$, and would give a cross at the column
containing its empty cell, since every column has a heavy cell.
Thus every row of size at least $4$ is full.
Since at least two rows have size at least $5$, at most one row is not full,
so some column $Y_t$ has no empty cell.
Relabel the rows so that $Z_{1t}$ is heavy and $X_2$ is another row of size at least $5$.
If $X_2$ has a cell of size at least $2$ outside $Y_t$, there is a cross at $Y_t$.
Otherwise, its three cells outside $Y_t$ are singletons, so $|Z_{2t}|\ge2$.
Choose $u\ne t$ such that $Z_{3u}$ is nonempty.
Column $Y_u$ has a heavy cell in $X_1$ or $X_3$, giving a cross at $Y_u$ with $X_2$.

When some column has degree $1$, after relabeling, say $Y_1\subseteq X_1$,
where $X_2$ and $X_3$ meet all of $Y_2,Y_3,Y_4$. The absence of a cross forces
$X_1$ to be full and all cells in $X_2$ and $X_3$ to have size at most $4$.
Consequently every cell in $X_1$ is heavy. Use the exchange
of~\cite[Theorem~4.1, Case~3]{HLMSX}:
\[
 \mathcal X'=\{X_2\cup X_3\cup Y_2\cup Y_3,Y_1,Z_{14}\},\qquad
 \mathcal Y'=\{Z_{22},Z_{23},Z_{32}\cup Z_{33},X_1\cup Y_4\}.
\]
The parts $Y_1$ and $Z_{14}$ ensure that $\mathcal X'$ is $5$-non-trivial.
The unions $Z_{32}\cup Z_{33}$ and $X_1\cup Y_4$ make $\mathcal Y'$
non-trivial, but $|Z_{22}|\le4$, a contradiction.

\emph{Case 2: exactly one row has degree $1$.}
After relabeling, say $X_1\subseteq Y_1$, where each of $X_2$ and $X_3$ meets
at least two of $Y_2,Y_3,Y_4$. We adapt the exchanges
of~\cite[Theorem~4.2]{HLMSX}.

When either $Z_{21}$ or $Z_{31}$ is empty, after relabeling, say $Z_{31}=\varnothing$.
Choose $\{j,k,l\}=\{2,3,4\}$ with $Z_{2j}\ne\varnothing$ so that one of
$X_1,Z_{3k},Z_{3l}$ has size at least $2$.
Such a choice exists: if $|X_1|=1$, then $|X_3|\ge5$, so some cell of $X_3$
outside $Y_1$ has size at least $2$, while $X_2$ has degree at least $3$.
Use the exchange
\[
 \mathcal X'=\{X_3\cup Y_k\cup Y_l,Y_1,Z_{2j}\},\qquad
 \mathcal Y'=\{X_2\cup Y_j,X_1,Z_{3k},Z_{3l}\}.
\]
The parts $X_3\cup Y_k\cup Y_l$ and $Y_1$ ensure that $\mathcal X'$ is
$5$-non-trivial. Together with $X_2\cup Y_j$, the part among
$X_1,Z_{3k},Z_{3l}$ of size at least $2$ makes $\mathcal Y'$ non-trivial. The
rows $X_1$ and $X_2$ are noncrossing with $\mathcal Y'$, giving compatibility.

When $Z_{21},Z_{31}\ne\varnothing$, relabel $X_2$ and $X_3$ and choose
$\{j,k,l\}=\{2,3,4\}$ so that the cells $Z_{2j},Z_{2k},Z_{3l}$ are nonempty,
one of $Z_{2j},Z_{2k}$ is heavy, and one of $X_1,Z_{31},Z_{3l}$ has size at
least $2$. Such a choice exists: each of $Y_2,Y_3,Y_4$ has a heavy cell in
$X_2$ or $X_3$, so after relabeling $X_2$ and $X_3$ the row $X_2$ has at least
two heavy cells $Z_{2j}$ and $Z_{2k}$ among $Y_2,Y_3,Y_4$. If $Z_{3l}$ is
heavy, all conditions hold. Otherwise $Z_{2l}$ is also heavy, and relabeling
$j,k,l$ so that $|Z_{3l}|\ge\max(|Z_{3j}|,|Z_{3k}|)$ ensures that
$\max(|Z_{31}|,|Z_{3l}|)\ge2$ whenever $|X_1|=1$. Use the exchange
\[
 \mathcal X'=\{X_3\cup Y_1\cup Y_l,Z_{2j},Z_{2k}\},\qquad
 \mathcal Y'=\{X_2\cup Y_j\cup Y_k,X_1,Z_{31},Z_{3l}\}.
\]
Then $\mathcal X'$ is $5$-non-trivial, $\mathcal Y'$ is non-trivial, and $X_1$
and $X_2$ are noncrossing with $\mathcal Y'$.

For the algorithm, let $n=|V|$. If $n<6$, return infeasible, and if
$6\le n\le51$, solve by enumeration. Otherwise enumerate each nonempty $S$ of
size at most $12$, combining $\{S\}$ with a minimum non-trivial 3-partition of
$f_{\setminus S}$ when $|S|=1$ and with a minimum 3-partition of
$f_{\setminus S}$ when $2\le|S|\le12$ (\cref{sec:primitives}). Since
$n-|S|\ge40$ when $|S|\ge2$, every resulting 4-partition is non-trivial. Also
compute a minimum $5$-non-trivial 3-partition $\mathcal X=\{X_1,X_2,X_3\}$
(\cref{sec:primitives}), recurse on $f_{/\{X_j,X_k\}}$ for each pair
$\{X_j,X_k\}\subseteq\mathcal X$, lift each returned 4-partition to $V$, and
return a candidate of minimum value.

If every minimum non-trivial 4-partition is $13$-size, some such $\mathcal Y$
satisfies $\mathcal X\compat\mathcal Y$. For the two parts
$X_j,X_k$ of $\mathcal X$ noncrossing with $\mathcal Y$, contracting $X_j$ and
$X_k$ changes at most two parts of $\mathcal Y$ and leaves at least two parts
of size at least $13$, so $\mathcal Y$ corresponds to a non-trivial 4-partition
of $f_{/\{X_j,X_k\}}$ of the same value. Conversely, every non-trivial
4-partition of $f_{/\{X_j,X_k\}}$ lifts to a non-trivial 4-partition of $V$.
For $a_i=|X_i|$, each recursive call has size $a_i+2<n$, and the measure $m-3$
on instances of size $m\ge6$ satisfies $\sum_{i=1}^3(a_i-1)=n-3$ with
$a_i-1\ge0$, at least two positive terms, and immediate termination when
$a_i+2<6$. Thus the recursion tree has $O(n)$ nodes, each taking polynomial
time.
\end{proof}


\section{Configurations of 4- and 5-Partitions}

Throughout this section, consider a bipartite graph on rows $X_1,\ldots,X_4$
and columns $Y_1,\ldots,Y_5$.
Call a row of degree $1$ a \emph{leaf row} and its unique incident edge a
\emph{leaf edge}.
Let $F=\{1,2,3,4,5\}$, and write
$N(X_i)\subseteq F$ for the set of indices $j$ such that $X_iY_j$ is an edge.
A \emph{connected $4/5$ split} is a partition of the vertices into sets
$U,W$ of sizes $4$ and $5$, respectively, each inducing a connected
subgraph.
When the graph is the configuration graph $G(M)$ of a 4-partition
$\mathcal X$ and a 5-partition $\mathcal Y$, applying \cref{lem:uncrossing}
to a connected $4/5$ split $(U,W)$ gives partitions
\[
 \mathcal X'=\{W^*\}\cup\mathcal L_U,\qquad
 \mathcal Y'=\{U^*\}\cup\mathcal L_W,\qquad
 f(\mathcal X')+f(\mathcal Y')\le f(\mathcal X)+f(\mathcal Y),
\]
where $\mathcal L_U$ and $\mathcal L_W$ have three and four parts,
respectively. If a cell internal to $U$ has size at least $2$, or if
$G(M)[U]$ contains a 4-cycle, then some part of $\mathcal L_U$ has size at
least $2$: in the latter case at least four nonempty cells internal to $U$
are distributed among three parts.
Together with a column of size at least $5$ in $W$, this makes
$\mathcal X'$ non-trivial. Placing any leaf rows belonging to $W$ last in the
vertex ordering in the proof of \cref{lem:uncrossing} retains each such leaf
row as a part of $\mathcal L_W$.

\begin{lemma}\label{lem:graph}
Let $G$ be a connected bipartite graph with rows
$X_1,\ldots,X_4$ and columns $Y_1,\ldots,Y_5$.
Assume that no row has degree $2$.
\begin{enumerate}[label=(\roman*),leftmargin=*]
\item Suppose that $G$ has at most one leaf row.
Then at least one of the following holds:
\begin{enumerate}[label=(\alph*),leftmargin=*]
\item Some column has, for each incident edge, a connected $4/5$ split $(U,W)$
in which $U$ contains both endpoints of the edge and exactly two rows.
\item There is a connected $4/5$ split $(U,W)$ such that $G[U]$ contains a
$4$-cycle.
\end{enumerate}
\item Suppose that $G$ has exactly two leaf rows with distinct neighbors
and is not, up to relabeling, the graph with leaf edges
$X_1Y_1$ and $X_2Y_2$ and neighborhoods
$N(X_3)=\{3,4,5\}$ and $N(X_4)=F$.
Then some column has, for each incident edge, a connected $4/5$ split $(U,W)$
in which $U$ contains both endpoints of the edge and one or two rows.
\end{enumerate}
In (i)(a) and (ii), the split may depend on the edge.
\end{lemma}

\begin{proof}
When $G$ has no leaf rows, every row has degree
at least $3$. We find a connected $4/5$ split $(U,W)$ such that
$G[U]$ contains a 4-cycle.
Pass to a spanning subgraph obtained by retaining three edges at each row
while keeping all columns covered:
if some column $Y_j$ is uncovered, replace a retained edge $X_iY_k$
by an unretained edge $X_iY_j$, where $Y_k$ is also covered by
another row, as any two rows cover at least one common column.
Let $K_i=F\setminus N(X_i)$ be the missing pair of $X_i$ in this subgraph.
We need a 2-element set $A$ of row indices, its complement
$B=\{1,2,3,4\}\setminus A$, and a 2-element set $L$ of column indices such
that
\[
 L\subseteq F\setminus\bigcup_{i\in A}K_i,
 \qquad \bigcap_{i\in B}K_i\subseteq L,
 \qquad F\setminus\bigcup_{i\in B}K_i\nsubseteq L.
\]
The first condition makes
$U=\{X_i:i\in A\}\cup\{Y_j:j\in L\}$ a 4-cycle.
The last two conditions ensure that the rows indexed by $B$
cover all columns outside $U$ and share at least one such column.
Thus the complement $W$ is connected.
When two missing pairs coincide, choose their row indices for $A$.
The rows indexed by $A$ cover the same three columns, including
those indexed by $J=\bigcap_{i\in B}K_i$.
If $|J|=2$, take $L=J$. If $|J|=1$, add any
other common neighbor index of the rows indexed by $A$.
If $J=\varnothing$, avoid the unique common neighbor
index of the rows indexed by $B$.
These choices satisfy all three conditions.
When the four missing pairs are distinct,
view them as four edges on $F$.
The star with four edges is excluded because every column is covered.
With $ij$ denoting $\{i,j\}$, each row of the table below satisfies the three
conditions, covering the remaining cases up to relabeling.
\begin{center}
\small
\begingroup
\setlength{\tabcolsep}{6pt}
\begin{tabular}{@{}lll@{}}
\toprule
Missing pairs & Missing pairs indexed by $A$ & $L$\\
\midrule
$12,23,34,45$ & $12,23$ & $4,5$\\
$12,13,14,45$ & $12,13$ & $4,5$\\
$12,23,34,14$ & $12,23$ & $4,5$\\
$12,23,13,14$ & $12,14$ & $3,5$\\
$12,23,13,45$ & $12,23$ & $4,5$\\
\bottomrule
\end{tabular}
\endgroup
\end{center}

When $G$ has a single leaf row $X_1$ adjacent to $Y_1$, let
$T=F\setminus\{1\}$ and write $T_i=N(X_i)\cap T$ for $i=2,3,4$.
Since $X_2,X_3,X_4$ each have degree at least $3$, we have $|T_i|\ge2$. Consider an arbitrary non-leaf edge at
$Y_1$, say $X_2Y_1$. For $t\in T_2$, try
$U=\{X_1,X_2,Y_1,Y_t\}$ and
$W=\{X_3,X_4\}\cup\{Y_j:j\in T\setminus\{t\}\}$.
If $|T_3\cup T_4|=3$, take the unique index $t\in T\setminus(T_3\cup T_4)$,
which lies in $T_2$ because $Y_t$ is covered.
If $T_3\cup T_4=T$ and $T_3\cap T_4\ne\varnothing$, choose $t\in T_2$
avoiding the unique element of $T_3\cap T_4$ when $|T_3\cap T_4|=1$.
Connectedness gives at least one non-leaf edge at $Y_1$, and any such split
also covers $X_1Y_1$, so $Y_1$ is the required column whenever this
construction works for every non-leaf edge at $Y_1$. Otherwise, assume that
it fails for $X_2Y_1$. Then $T_3$ and $T_4$ are equal or disjoint 2-element
sets. If $T_3=T_4$, then $G[U]$ is a 4-cycle for $U=\{X_3,X_4\}\cup\{Y_j:j\in T_3\}$,
and its complement is connected because $X_2$ is adjacent to $Y_1$ and to both
columns indexed by $T\setminus T_3$.
If $T_3\cap T_4=\varnothing$, both $X_3$ and $X_4$ are adjacent to $Y_1$.
Interchange $X_3$ and $X_4$ if needed, choose
$c\in T_2\cap T_3$, and use
$U=\{X_2,X_3\}\cup\{Y_j:j\in T_3\}$.
Its complement is connected through $X_4$ and $Y_1$.
The only edges incident with $Y_c$ are $X_2Y_c$ and $X_3Y_c$, both inside
$G[U]$. Thus $Y_c$ is the required column.

When the leaf edges are $X_1Y_1$ and $X_2Y_2$, let $T=\{3,4,5\}$.
Write $H_i=N(X_i)\cap\{1,2\}$ and $T_i=N(X_i)\cap T$ for $i=3,4$.
Connectedness gives $H_3\cup H_4=\{1,2\}$ and $T_3\cup T_4=T$.
\cref{fig:graph-splits} gives the two split templates.
Write $I_3(t)$ for template (a), and $II_3$ for template (b).
The templates with subscript $4$ are obtained by interchanging $X_3,X_4$ and, for template (a), also
$X_1,X_2$ and $Y_1,Y_2$.

\begin{figure}[htbp]
\centering
\begingroup
\setlength{\tabcolsep}{4pt}
\renewcommand{\arraystretch}{1.05}
\begin{subfigure}[b]{.32\textwidth}
\centering
\begin{tabular}{c|cc|ccc|}
\multicolumn{1}{c}{} & \multicolumn{1}{c}{$Y_1$} & \multicolumn{1}{c}{$Y_t$}
  & \multicolumn{1}{c}{$Y_2$} & \multicolumn{1}{c}{$Y_u$} & \multicolumn{1}{c}{$Y_v$} \\
\hhline{~|-----|}
$X_1$ & \cellcolor{\blue}$1^+$ & \cellcolor{\blue}$0$ & $0$ & $0$ & $0$ \\
$X_3$ & \cellcolor{\blue}$1^+$ & \cellcolor{\blue}$1^+$ & $?$ & $?$ & $?$ \\
\hhline{~|-----|}
$X_2$ & $0$ & $0$ & \cellcolor{\green}$1^+$ & \cellcolor{\green}$0$ & \cellcolor{\green}$0$ \\
$X_4$ & $?$ & $?$ & \cellcolor{\green}$1^+$ & \cellcolor{\green}$1^+$ & \cellcolor{\green}$1^+$ \\
\hhline{~|-----|}
\end{tabular}
\caption{$U=\{X_1,X_3,Y_1,Y_t\}$}
\end{subfigure}\hfill
\begin{subfigure}[b]{.32\textwidth}
\centering
\begin{tabular}{c|ccc|cc|}
\multicolumn{1}{c}{} & \multicolumn{1}{c}{$Y_3$} & \multicolumn{1}{c}{$Y_4$}
  & \multicolumn{1}{c}{$Y_5$} & \multicolumn{1}{c}{$Y_1$} & \multicolumn{1}{c}{$Y_2$} \\
\hhline{~|-----|}
$X_3$ & \cellcolor{\blue}$1^+$ & \cellcolor{\blue}$1^+$ & \cellcolor{\blue}$1^+$ & $?$ & $?$ \\
\hhline{~|-----|}
$X_1$ & $0$ & $0$ & $0$ & \cellcolor{\green}$1^+$ & \cellcolor{\green}$0$ \\
$X_2$ & $0$ & $0$ & $0$ & \cellcolor{\green}$0$ & \cellcolor{\green}$1^+$ \\
$X_4$ & $?$ & $?$ & $?$ & \cellcolor{\green}$1^+$ & \cellcolor{\green}$1^+$ \\
\hhline{~|-----|}
\end{tabular}
\caption{$U=\{X_3,Y_3,Y_4,Y_5\}$}
\end{subfigure}\hfill
\begin{subfigure}[b]{.32\textwidth}
\centering
\begin{tabular}{c|ccccc|}
\multicolumn{1}{c}{} & \multicolumn{1}{c}{$Y_1$} & \multicolumn{1}{c}{$Y_2$}
  & \multicolumn{1}{c}{$Y_3$} & \multicolumn{1}{c}{$Y_4$} & \multicolumn{1}{c}{$Y_5$} \\
\hhline{~|-----|}
$X_1$ & $1^+$ & $0$ & $0$ & $0$ & $0$ \\
$X_2$ & $0$ & $1^+$ & $0$ & $0$ & $0$ \\
$X_3$ & $0$ & $0$ & $1^+$ & $1^+$ & $1^+$ \\
$X_4$ & $1^+$ & $1^+$ & $1^+$ & $1^+$ & $1^+$ \\
\hhline{~|-----|}
\end{tabular}
\caption{The exception}
\end{subfigure}
\endgroup
\caption{The two-leaf case of \cref{lem:graph}.
In (a), $\{t,u,v\}=\{3,4,5\}$. The blue and green blocks mark the row-column
pairs with both vertices in $U$ and in $W$, respectively, and the entries
marked $1^+$ ensure that $G[U]$ and $G[W]$ are connected. In (c), $G$ is the
excluded graph in \cref{lem:graph}(ii). The symbols $1^+$ and $?$ have the
meanings in \cref{fig:cross-exchange}.}
\label{fig:graph-splits}
\end{figure}


Up to these interchanges, $(H_3,H_4)$ has the four types below. Each row gives
a column index $t$ and splits covering all edges at $Y_t$, allowing different
splits per edge.
\begin{center}
\small
\begingroup
\setlength{\tabcolsep}{6pt}
\begin{tabular}{@{}llll@{}}
\toprule
$(H_3,H_4)$ & $(T_3,T_4)$ & Column index $t$ & Splits\\
\midrule
$(\varnothing,\{1,2\})$ & $T_4\ne T$ & $t\in T\setminus T_4$ & $II_3$\\
 & $T_4=T$ & \multicolumn{2}{l}{the exception in \cref{fig:graph-splits}(c)}\\
$(\{1\},\{2\})$ & $T_3\ne T$ & $t\in T\setminus T_3$ & $I_4(t)$\\
 & $T_3=T_4=T$ & any $t\in T$ & $I_3(t),I_4(t)$\\
$(\{1\},\{1,2\})$ & $T_3\ne T$ & $t\in T\setminus T_3$ & $I_4(t)$\\
 & $T_3=T$ & $t\in T_4$ & $II_3,I_4(t)$\\
$(\{1,2\},\{1,2\})$ & $T_3=T$ & $t\in T_4$ & $II_3,I_4(t)$\\
 & $T_3,T_4\ne T$ and $|T_3|=2$ & $t\in T\setminus T_3$ & $I_4(t)$\\
\bottomrule
\end{tabular}
\endgroup
\end{center}
In the second type, interchange $(X_1,X_3,Y_1)$ with
$(X_2,X_4,Y_2)$ if needed. In the fourth, interchange $X_3,X_4$ if needed.
A proper subset $T_3\subsetneq T$ used in the table has size $2$, so
$T\setminus\{t\}=T_3$ makes $I_4(t)$ valid.
In the last two types, when $T_3=T$, template $II_3$ covers $X_3Y_t$
and $I_4(t)$ covers $X_4Y_t$ at any $t\in T_4$.
The first type has $T_3=T$ by the degree bound, giving either $II_3$ or
the stated exception.
\end{proof}

\begin{lemma}\label{lem:canonical-short}
Let $f:2^V\to\mathbb R$ be submodular, let $\mathcal X$ be a minimum
non-trivial $4$-partition of $V$, and let $\mathcal Y$ be the unique minimum
$5$-partition of $V$. If $\mathcal Y$ is $5$-size and $\mathcal X$ is
incompatible with $\mathcal Y$, their configuration has the nonempty cells in
\cref{fig:canonical-configuration}, up to relabeling.
\end{lemma}
\begin{figure}[htbp]
\centering
\begingroup
\setlength{\tabcolsep}{9pt}
\renewcommand{\arraystretch}{1.05}
\begin{tabular}{c|ccccc|}
\multicolumn{1}{c}{} & \multicolumn{1}{c}{$Y_1$} & \multicolumn{1}{c}{$Y_2$}
  & \multicolumn{1}{c}{$Y_3$} & \multicolumn{1}{c}{$Y_4$} & \multicolumn{1}{c}{$Y_5$} \\
\hhline{~|-----|}
$X_1$ & $Z_{11}$ & $0$      & $0$      & $0$      & $0$ \\
$X_2$ & $Z_{21}$ & $0$      & $0$      & $0$      & $0$ \\
$X_3$ & $Z_{31}$ & $Z_{32}$ & $Z_{33}$ & $0$      & $0$ \\
$X_4$ & $Z_{41}$ & $0$      & $0$      & $Z_{44}$ & $Z_{45}$ \\
\hhline{~|-----|}
\end{tabular}
\endgroup
\caption{The configuration in \cref{lem:canonical-short}.
Rows are parts of $\mathcal X$ and columns are parts of $\mathcal Y$. Named
cells are nonempty, and $0$ denotes an empty cell. Thus
$X_3=Z_{31}\cup Y_2\cup Y_3$ and $X_4=Z_{41}\cup Y_4\cup Y_5$, with
$Y_1=X_1\cup X_2\cup Z_{31}\cup Z_{41}$. This is the obstruction
in~\cite[Section~7]{HLMSX}.}
\label{fig:canonical-configuration}
\end{figure}


\begin{proof}
If a row $X_i$ has degree $2$, apply \cref{lem:uncrossing} to the vertex set
consisting of $X_i$ and its two neighboring columns, which induces a connected
subgraph.
The new 5-partition consists of the two nonempty cells of $X_i$ and
the other three rows, while the new 4-partition consists of the
resulting union and the three untouched columns. The latter is non-trivial.
Since $\mathcal X$ is minimum among non-trivial 4-partitions, the new
5-partition is also minimum and $\mathcal X$ is compatible with it,
contradicting the uniqueness of $\mathcal Y$.
This is the degree-$2$ exchange of~\cite[Lemma~4.2]{HLMSX}, with one additional
untouched row and column. Thus every row has degree $1$ or at least $3$,
and at most two rows are leaf rows.

We claim that two leaf rows share a column. Suppose otherwise. If $G(M)$
is connected and is not the exception in \cref{lem:graph}(ii), each column
has size at least $5$ and therefore contains a cell of size at least $2$.
Applying \cref{lem:graph} (at such a cell in cases (i)(a) and (ii)) and
\cref{lem:uncrossing} yields a non-trivial 4-partition $\mathcal X'$ and a
minimum 5-partition $\mathcal Y'=\{U^*\}\cup\mathcal L_W$. Either $U$
contains two rows, so the part $U^*$ of $\mathcal Y'$ contains two rows of
$\mathcal X$, or $U$ contains one row and $W$ contains both leaf rows, which
remain separate parts of $\mathcal L_W$ while the row in $U$ lies in $U^*$,
giving $\mathcal X\compat\mathcal Y'$. Since no two leaf rows share a column,
no part of $\mathcal Y$ contains two rows of $\mathcal X$, and $\mathcal X$ is
incompatible with $\mathcal Y$. Thus $\mathcal Y'\ne\mathcal Y$ in both cases,
contradicting the uniqueness of $\mathcal Y$.
For the exception, with $X_3=Z_{33}\cup Z_{34}\cup Z_{35}$ and
$X_4=Z_{41}\cup\cdots\cup Z_{45}$, uncrossing $X_3$ successively with $Y_3$
and $Y_5$, and $X_4$ with $Y_4$, gives
\[
 \mathcal X'=\{Y_1,Y_2,X_3\cup Z_{43}\cup Z_{45},Z_{44}\},\qquad
 \mathcal Y'=\{X_1,X_2,Z_{33},X_4\cup Z_{34},Z_{35}\},
\]
again with non-trivial $\mathcal X'$ and $\mathcal X\compat\mathcal Y'$.

If $G(M)$ is disconnected, all rows of degree at least $3$ lie in the same
component, since any such row has at least three neighbors among the five
columns. Unless two leaf rows
share a column, another component is an isolated edge, say $X_1=Y_j$.
For $f_{\setminus X_1}$, the partition $\mathcal Y\setminus\{Y_j\}$ is the
unique minimum 4-partition and is $5$-size. The restricted ground set has at
least $20$ elements. Every 3-partition of this ground set
has a part of size at least $2$, so adjoining $X_1$ gives a non-trivial
4-partition.
Thus the three remaining rows of $\mathcal X$ form a minimum 3-partition of
$f_{\setminus X_1}$. Either they are non-trivial, or their two singletons
together with $X_1$ already make $\mathcal X$ compatible with $\mathcal Y$.
In the former case, \cite[Theorem~4.3]{HLMSX} implies that two of the
remaining rows are noncrossing with $\mathcal Y\setminus\{Y_j\}$, hence with
$\mathcal Y$, contradicting incompatibility. This proves the claim.

Relabel so that the leaf rows are $X_1$ and $X_2$, both adjacent to $Y_1$.
The rows $X_3$ and $X_4$ both have degree at least $3$, and
$Y_t=Z_{3t}\cup Z_{4t}$ for every $t\in F\setminus\{1\}$.
For any 2-element set $S\subseteq N(X_3)$ such that $T=N(X_3)\setminus S$
satisfies $1\notin T$ and $T\subseteq N(X_4)$, applying
\cref{lem:uncrossing} to $\{X_3\}\cup\{Y_j:j\in S\}$ and to
$\{X_4\}\cup\{Y_t:t\in T\}$ gives
\[
\begin{split}
\mathcal X'&=\{X_3\cup\bigcup_{j\in S}Y_j\}
 \cup\{Z_{4t}:t\in T\}\cup\{Y_j:j\in F\setminus N(X_3)\},\\
\mathcal Y'&=\{X_1,X_2\}\cup\{Z_{3j}:j\in S\}
 \cup\{X_4\cup\bigcup_{t\in T}Y_t\},
\end{split}
\]
which have four and five parts, respectively. Whenever $\mathcal X'$ is
non-trivial, $\mathcal Y'$ is also a minimum 5-partition and satisfies
$\mathcal X\compat\mathcal Y'$, contradicting the uniqueness of $\mathcal Y$.
If $1\notin N(X_3)$, choose $T$ to consist of one index in
$N(X_3)\cap(N(X_4)\setminus\{1\})$ when $|N(X_3)|=3$, or two indices in
$N(X_4)\setminus\{1\}$ when $N(X_3)=F\setminus\{1\}$. In both cases the
untouched column $Y_1$ makes $\mathcal X'$ non-trivial.
By symmetry, $1\in N(X_3)\cap N(X_4)$. The sets
$A_3=N(X_3)\setminus\{1\}$ and $A_4=N(X_4)\setminus\{1\}$ each have size at
least $2$ and satisfy $A_3\cup A_4=F\setminus\{1\}$. Unless $|A_3|=|A_4|=2$,
relabel $X_3,X_4$ so that $|A_3|\le|A_4|$, giving $|A_4|\ge3$ and
$|A_3\setminus A_4|\le1$, and choose $c\in A_3$ with
$A_3\setminus A_4\subseteq\{c\}$. Setting $S=\{1,c\}$ gives $1\notin T$ and
$T\subseteq A_4$. If $|A_3|<4$, an untouched column in $F\setminus N(X_3)$
makes $\mathcal X'$ non-trivial. If $|A_3|=|A_4|=4$, each column $Y_t$ with
$t\in F\setminus\{1\}$ has size at least $5$, so relabeling $X_3,X_4$ gives
$|Z_{4t}|\ge2$ for some $t\in F\setminus\{1\}$, and choosing
$c\in F\setminus\{1,t\}$ gives $t\in T$ so that $X_3\cup Y_1\cup Y_c$ and
$Z_{4t}$ make $\mathcal X'$ non-trivial. Hence $|A_3|=|A_4|=2$, and
$A_3\cup A_4=F\setminus\{1\}$ makes $A_3$ and $A_4$ disjoint, as asserted.
\end{proof}

\begin{lemma}\label{lem:noncrossing}
Let $f:2^V\to\mathbb R$ be submodular, let $\mathcal X$ be a minimum
non-trivial $4$-partition of $V$, and let $\mathcal Y$ be the unique minimum
$5$-partition of $V$. Suppose $\mathcal Y$ is $5$-size and $\mathcal X$ is
incompatible with $\mathcal Y$, and let $\mathcal Z=\{P,Q,R\}$ minimize $f$
over $\mathcal F_{\mathcal X}$ with $P\cup Q$ in one part of $\mathcal X$. Then
$P$ and $Q$ are noncrossing with $\mathcal Y$.
\end{lemma}
\begin{proof}
Suppose otherwise and, by symmetry, assume that $P$ crosses $\mathcal Y$.
By \cref{lem:canonical-short}, $X_1\cup X_2\subseteq Y_1$,
so $P\cup Q$ is contained in $X_3$ or $X_4$.
Interchanging $X_3$ and $X_4$ and the column pairs
$(Y_2,Y_3)$ and $(Y_4,Y_5)$ if needed, assume $P\cup Q\subseteq X_3$, so that
$Y_2=Z_{32}$ and $Y_3=Z_{33}$, while $Y_4\cup Y_5\subseteq X_4$. Let
$P_i=P\cap Z_{3i}=P\cap Y_i$ for $i=1,2,3$. Since $P$ crosses $\mathcal Y$,
at least two of the three sets $P_i$ are nonempty.

If all three $P_i$ are nonempty, \cref{lem:uncrossing} gives
$f(P)+f(Y_1)+f(Y_2)+f(Y_3)\ge f(X_3\cup Y_1)+\sum_{i=1}^3 f(P_i)$,
so $\mathcal Z'=\{X_3\cup Y_1,Y_4,Y_5\}$ and
$\mathcal Y'=\{P_1,P_2,P_3,Q,R\}$ satisfy
$f(\mathcal Z')+f(\mathcal Y')\le f(\mathcal Z)+f(\mathcal Y)$.
Here $\mathcal Z'\in\mathcal F_{\mathcal X}$ since $Y_4,Y_5\subseteq X_4$ have
size at least $5$, making $\mathcal Y'$ a minimum 5-partition with
$\varnothing\subsetneq P_1\subsetneq Y_1$, a contradiction.

Otherwise, exactly two of $P_1,P_2,P_3$ are nonempty, say $P_i$ and $P_j$, with
$\{i,j,k\}=\{1,2,3\}$. Consider
\[
 \mathcal X'=\{P_i,P_j,Q,R\},\quad
 \mathcal Y'=\{X_1,X_2,Z_{3i}\cup Z_{3j},Z_{3k},X_4\},\quad
 \mathcal Z'=\{X_3\cup Y_1,Y_4,Y_5\}.
\]
The 4-partition $\mathcal X'$ is non-trivial because $\mathcal Z$ has at
least two parts of size at least $3$, and splitting $P$ into $P_i$ and $P_j$
produces a part of size at least $2$ whenever $|P|\ge3$. Moreover,
$\mathcal Y'$ is a 5-partition and $\mathcal Z'\in\mathcal F_{\mathcal X}$.
Submodularity gives
\[
\begin{split}
f(X_3)+f(Y_1)&\ge f(Z_{31})+f(X_3\cup Y_1),\\
f(Z_{3i})+f(Z_{3j})+f(P)&\ge f(P_i)+f(P_j)+f(Z_{3i}\cup Z_{3j}),
\end{split}
\]
where the second inequality applies submodularity to $Z_{3i}$ and
$P=P_i\cup P_j$ and then to $Z_{3i}\cup P_j$ and $Z_{3j}$. Adding the remaining
part values gives
$f(\mathcal X')+f(\mathcal Y')+f(\mathcal Z')\le f(\mathcal X)+f(\mathcal Y)+f(\mathcal Z)$,
using $Y_2=Z_{32}$ and $Y_3=Z_{33}$ together with
$f(Z_{3i})+f(Z_{3j})+f(Z_{3k})=f(Z_{31})+f(Z_{32})+f(Z_{33})$.
Since the partitions $\mathcal X,\mathcal Y,\mathcal Z$ are optimal in their
respective families, $\mathcal Y'$ is also a minimum 5-partition.
But $X_1$ and $X_2$ are separate in $\mathcal Y'$ and together in $\mathcal Y$, contradicting the uniqueness of $\mathcal Y$.
\end{proof}


\section{The Algorithm}

For a submodular function $g$ on a current ground set $W$ with $|W|\ge5$, let
$\mathcal B(g)$ be a minimum 5-partition of $W$ among those having a part of
size at most $4$. Enumerate each nonempty $S\subseteq W$ with
$|S|\le\min(4,|W|-4)$ and combine $\{S\}$ with a minimum 4-partition of
$g_{\setminus S}$. This computes $\mathcal B(g)$ in polynomial time
by \cref{sec:primitives} (see~\cite[Theorem~6.1]{HLMSX}), and for
$5\le|W|\le24$ it returns a minimum 5-partition of $g$.

Given $f:2^{V_0}\to\mathbb R$, pass to the lexicographic pair $(f,\psi)$ from
\cref{sec:primitives}, which we still write as $f$. By \cref{sec:primitives},
$\mathcal B(g)$ and \cref{lem:auxiliary-three,lem:nontrivial-four} apply
directly to $f$ and its contractions, and every instance has a unique minimum
5-partition. \cref{alg:five-partition} gives the recursive procedure
$\textsc{Min5Partition}(f)$, where $V$ is the ground set of the current call
and every $\operatorname{lift}(\cdot)$ is to $V$.

\begin{figure}[htbp]
\centering
\small
\begin{algorithm}
\textul{$\textsc{Min5Partition}(f\text{ on }V)$:}\+
\\ if $|V|<5$, return infeasible
\\ if $|V|\le24$, return $\mathcal B(f)$
\\ $\mathcal X\gets\arg\min\{f(\mathcal P):\mathcal P\text{ is a non-trivial 4-partition}\}$
\\ $\mathcal C\gets\varnothing$
\\ for $X_i\in\mathcal X$ with $|X_i|\ge2$\+
\\ add $\operatorname{lift}(\textsc{Min5Partition}(f_{/(\mathcal X\setminus\{X_i\})}))$ to $\mathcal C$\-
\\ $\mathcal U\gets\arg\min_{\mathcal P\in\mathcal C}f(\mathcal P)$, and $g\gets f$
\\ repeat\+
\\ $\mathcal U\gets\arg\min\{f(\mathcal P):\mathcal P\in\{\mathcal U,\operatorname{lift}(\mathcal B(g))\}\}$
\\ if the ground set of $g$ has size at most $24$ or $\mathcal X$ is trivial, return $\mathcal U$
\\ $\{P,Q,R\}\gets\arg\min_{\mathcal Z\in\mathcal F_{\mathcal X}}g(\mathcal Z)$,
 with $P\cup Q\subseteq X_i\in\mathcal X$
\\ $g\gets g_{/\{P,Q\}}$, and replace $X_i$ in $\mathcal X$ by $(X_i\setminus(P\cup Q))\cup\{u_P,u_Q\}$\-\-
\end{algorithm}
\caption{Compute a minimum 5-partition.}
\label{alg:five-partition}
\end{figure}

Whenever an iteration reaches the minimization over $\mathcal F_{\mathcal X}$,
the current ground set of $g$ has at least $25$ elements, so some part of
$\mathcal X$ has at least $7$ elements. Taking three of them as $P$, one as
$Q$, and the rest of the ground set as $R$ shows that
$\mathcal F_{\mathcal X}$ is nonempty. At least one of $P$ and $Q$ has size at
least $3$, so contracting $P$ and $Q$ reduces the ground-set size of $g$ by at
least $2$. Moreover, the updated 4-partition $\mathcal X$ has $\{u_P,u_Q\}$ in
one part and three other nonempty parts, so the ground set of $g$ always has
at least $5$ elements.

\begin{proof}[\cref{thm:main}]
Run $\textsc{Min5Partition}$ on $(f,\psi)$ (still written $f$) and induct on
$|V|$. For $|V|\le24$, the call returns $\mathcal B(f)$, which is optimal.
Assume $|V|\ge25$ and let $\mathcal Y^*$ be the unique minimum 5-partition of
$f$. Every candidate placed in $\mathcal C$ or compared with $\mathcal U$
lifts a 5-partition to $V$, so $f(\mathcal U)\ge f(\mathcal Y^*)$ throughout.
If $\mathcal X\compat\mathcal Y^*$, then for some $X_i\in\mathcal X$ the parts
in $\mathcal X\setminus\{X_i\}$ are noncrossing with $\mathcal Y^*$, so
$\mathcal Y^*$ lifts from a minimum 5-partition of
$f_{/(\mathcal X\setminus\{X_i\})}$. That contracted ground set has size
$|X_i|+3\ge5$, so $|X_i|\ge2$ and the recursive call for $X_i$ sets
$\mathcal U=\mathcal Y^*$ before the loop.

Suppose $\mathcal X$ is incompatible with $\mathcal Y^*$. At the start of each
iteration, either $\mathcal U=\mathcal Y^*$, or $g$ has a unique minimum
5-partition $\mathcal Y$ lifting to $\mathcal Y^*$ and $\mathcal X$ is a
minimum non-trivial 4-partition of $g$ incompatible with $\mathcal Y$. If
$\mathcal U\ne\mathcal Y^*$ after comparing with
$\operatorname{lift}(\mathcal B(g))$, then $\mathcal Y$ is $5$-size, $g$ has
at least $25$ elements, and $\mathcal X$ is non-trivial, so the iteration
continues. By \cref{lem:canonical-short,lem:noncrossing}, $P$ and $Q$ each
lie in a cell of $\mathcal X$ and $\mathcal Y$. Contracting $P$ and $Q$ maps
$\mathcal Y$ to a 5-partition $\mathcal Y'$ of $g_{/\{P,Q\}}$ that has the
same configuration graph with the updated $\mathcal X$, so the updated
$\mathcal X$ remains incompatible with $\mathcal Y'$ and its two crossing rows
still have degree $3$, keeping $\mathcal X$ non-trivial. Since lifting
preserves values, is injective, and maps non-trivial 4-partitions to
non-trivial 4-partitions, $\mathcal Y'$ is the unique minimum 5-partition of
$g_{/\{P,Q\}}$ and the updated $\mathcal X$ remains a minimum non-trivial
4-partition. Each iteration removes at least $2$ elements, so the loop
terminates and returns $\mathcal Y^*$.

For the running time, let $n=|V_0|$. On an $m$-element instance with $m\ge25$
and $a_i=|X_i|$, each recursive call has $a_i\ge2$, size $a_i+3<m$, and
positive measure $(a_i+3)-4=a_i-1$. Since $\sum_{i=1}^4(a_i-1)=m-4$ has at
least two positive terms, there are $O(n)$ invocations. Each invocation acts on
a contraction of $(f,\psi)$ on at most $n$ elements with second coordinates of
bit length polynomial in $n$ (\cref{sec:primitives}), calls
\cref{lem:nontrivial-four} once before the loop, and runs $O(n)$ iterations
that each compute $\mathcal B(g)$ and a minimum partition in
$\mathcal F_{\mathcal X}$ (\cref{lem:auxiliary-three}), so the algorithm runs
in polynomial time.
\end{proof}


\label{maintext-end}
\clearpage
\paragraph{Acknowledgments} GPT-6-Astra were used during the research for proof exploration and
checking, and for drafting and editing the manuscript. The authors are
responsible for the results and exposition.

\end{document}